\documentclass[usenatbib,usedcolumn,usegraphicx]{mn2e}  

\usepackage{times}

\usepackage[dvips]{color}

\newcommand{\apj}{ApJ}

\newcommand{\apjs}{ApJS}
\newcommand{\aj}{AJ}

\newcommand{\aap}{A\&A}

\title[Galaxy Number Counts and Implications for Strong Lensing]
{Galaxy Number Counts and Implications for Strong Lensing
}

\author[C. D. Fassnacht et al.]{
C. D. Fassnacht,$^1$, 
L. V. E. Koopmans,$^2$
and K.C. Wong$^{1,3}$ \\
$^1$Department of Physics, University of California Davis, 1 Shields Avenue,
   Davis, CA 95616, USA \\
$^2$Kapteyn Astronomical Institute, P.O. Box 800, 9700 AV Groningen, 
    Netherlands \\
$^3$Current Address:
Astronomy Department/Steward Observatory, 
University of Arizona, 
933 N. Cherry Ave.,
Tucson, AZ 85721, USA
}

\begin{document}

\maketitle
\begin{abstract}
We compare galaxy number counts in Advanced Camera for Surveys (ACS)
fields containing moderate-redshift ($0.2<z<1.0$) strong gravitational
lenses with those in two control samples: (1) the first square degree
of the COSMOS survey, comprising 259 ACS fields and (2) 20 ``pure
parallel'' fields randomly located on the sky.  Through a Bayesian
analysis we determine the expectation values ($\mu_0$) and confidence
levels of the underlying number counts for a range of apertures and
magnitude bins.  Our analysis has produced the following results:
(i)~We infer that our control samples are not consistent, with the
number counts in the COSMOS sample being significantly higher than in
the pure parallel sample for $21\leq F814W \leq 23$.  This result
matches those found in previous analyses of COSMOS data using
different techniques.  (ii)~We find that small-size apertures,
centered on strong lenses, are overdense 
compared with randomly placed apertures in the control samples, even
compared to the COSMOS sample.  Correcting for the local clustering of
elliptical galaxies, based on the average two-point correlation
function reduces this overdensity to the 1--2-$\sigma$ level.
Thus, the overdensity of galaxies seen along a typical line of sight
to a lens can be explained mostly by the natural clustering of galaxies,
rather than being due to lenses lying along otherwise biased lines of
sight.  
However, a larger sample of lenses is needed to see if the remaining
bias persists when the lens-field uncertainties are smaller.
(iii)~There is considerable scatter in the lines of sight to
{\it individual} lens systems, but quantities that are linearly
dependent on the external convergence (e.g., $H_0$) should become
unbiased if the extra galaxies that cause the bias can be accounted
for in the lens models either through direct modeling or via an
informed prior on the external convergence.  The number counts can
used to set such an informed prior.


\end{abstract}


\section{Introduction}
Strong gravitational lenses, where multiple images of the background
object are formed, are powerful probes of the distribution of mass in
the Universe.  The properties of the lensed images are, in principle,
sensitive only to the projected mass of the lensing object, with no
requirements that the mass be luminous or baryonic
\citep[e.g.,][]{saasfee}.  Most of the lensing signal comes from the
primary lensing object -- typically a massive early-type galaxy --
and, if the lensing object is a member of a galaxy cluster or group,
its immediate environment.

However, the distribution of large-scale structure (LSS) along the
line of sight to the lens system adds perturbations to the lensing
properties.  For example, simulations have shown that a non-negligible
fraction of lenses can only be produced by having multiple lens planes
along the line of sight \citep{multi_lens_planes, hilbert_ms1,
hilbert_ms2}.  Furthermore, the differences in light travel times
along the rays that form the multiple lensed images in a given system can be
affected at the level of a few percent \citep[e.g.,][]{seljaklss} or
up to $\sim$10\% \citep[e.g.,][]{barkanalss} by the distribution of
LSS along that particular line of sight.
These effects should be random for random lines of sight, so it
should be possible to reduce the LSS uncertainties and exploit the
power of gravitational lenses as cosmological tools by averaging over
many systems.  If, however, lenses lie along biased lines of sight, this
reduction will not occur and global parameters such as $H_0$
determined from large lens samples will be biased.

To date, most observational investigations of the effects of the
environment on strong lensing have focused on the local neighborhood
of the lens by searching for spectroscopic
evidence of galaxy groups and clusters that are physically associated
with the lensing galaxy
\citep[e.g.,][]{kundic1115,kundic1422,0712group,
  momcheva,1608groupdisc,auger1600,auger1520}.  Some of these
investigations
\citep[e.g.,][]{0712group,momcheva,1608groupdisc,auger1600} have also
found mass concentrations along the line of sight that are at
different redshifts than the lensing galaxy.  However, due to the
limitations imposed by requiring spectroscopic redshifts for their
analyses, the spectroscopic surveys are necessarily incomplete samples
of the line of sight.  Most also are also biased because they
preferentially target galaxies expected to be at the redshift of the
primary lens.  There have also been photometric studies of lens fields
in order to evaluate lens environments.  These photometric
investigations are the closest in concept to the analysis in this
paper.  However, they focused on either group finding via detection of
red sequences or spatial overdensities \citep{williams,faure_2004}, or
on describing the immediate environment of the lensing galaxy by using
galaxy colors to strongly favor galaxies likely to be at the redshift
of the lens \citep{auger_slacs,treu_slacs_environ}.  The
\citet{auger_slacs} work also evaluates the contributions by galaxies
along the full line of sight, and so has a component that is very
similar to the work presented in this paper.  However, it focuses
strictly on low redshift lenses ($z\sim 0.1 - 0.3$), while we study a
higher redshift sample ($0.2 < z < 1.0$).

In this paper, we investigate the question of whether lens lines of
sight are biased by comparing, through Bayesian and frequentist
statistics, the number counts of galaxies in fields containing
gravitational lenses with those obtained from two control samples.
The control samples are chosen to provide reasonable approximations to
typical lines of sight through the Universe.  All images were obtained
with the Advanced Camera for Surveys \citep[ACS;][]{acs1,acs2} aboard
the {\em Hubble Space Telescope} ({\em HST}).
The underlying idea is that lines of sight that are overdense in
galaxies are also overdense in mass, if the underlying redshift
distributions are roughly the same. Hence by simply counting galaxies,
one can make conservative statements about the lines of sight
towards lens galaxies that are not highly model-dependent.  
We discuss the sample selection and image processing in \S2, do
a frequentist analysis of the samples in \S3, develop and use a
Bayesian framework for comparing the samples in \S4, briefly
describe the number counts for individual lens systems (as opposed
to sample averages) in \S5, and interpret the results in \S6.

\section{Data Reduction}

In this section we briefly describe the data reduction and
catalogue extraction.

\subsection{Sample Definition and Data Processing}

The lens sample comprises 18 systems which were observed with ACS as
part of the CfA-Arizona Space Telescope Lens Survey (CASTLES; GO-9744;
PI Kochanek).  This sample was defined by taking the full list of lenses
observed with ACS as part of the CASTLES program (24 galaxies in all) and
only including systems that (1) had total exposure times $\ga$2000~s, (2) had
no extremely bright stars in the field, and (3) had  galactic
latitudes of $|b| > 10$.
Each system was observed through the F555W and F814W filters, but for
comparison with the control data sets, we only consider the F814W
data.  The typical total exposure times through the F814W filter were
$\sim$2000--3000~sec.  Details of the observations are given in
Table~1.  The pipeline-processed data were obtained from the
Multi-mission Archive at Space Telescope, and the individual exposures
were combined using the {\em multidrizzle} package
\citep{multidrizzle}. We also included in the lens sample deep ACS
images of B0218+357 (GO-9450; PI N.\ Jackson) and B1608+656 (GO-10158;
PI Fassnacht), with total F814W exposure times of 48,720 and
28,144~sec, respectively.  For easier comparison with the rest of the
lens sample we only used the data from the first four F814W exposures on
the B1608+656 field, with a combined integration time of 2528~sec.

The lens galaxies are at moderate redshifts, with a mean and RMS of
$\mu_z = 0.55$ and $\sigma_z = 0.22$, respectively.  This should be
compared to the environmental investigations of \citet{auger_slacs}
and \citet{treu_slacs_environ}, which analyzed the lower redshift
SLACS ($z\la 0.3$) sample \citep[e.g.,][]{slacsI,slacsV}.  The lens
galaxies can be further distinguished from the SLACS lenses in that
nearly all of them (17/20) are lensing active galactic nuclei (AGN)
rather than galaxies, and all but one of them were selected by
targeting the lensed source population rather than targeting the
likely lenses as was done in SLACS.  Most lens systems with existing
time delays have lens redshifts at $z>0.3$.  Thus, this sample, with
$z_{\rm lens} \sim 0.5$, and lensed sources that are expected to be
variable, is more likely to be representative of lens systems for
which time delays, and thus quantities such as $H_0$, can be measured.

\begin{table}
\caption{Lens Sample}
\label{tab_obsdata}
\begin{tabular}{lrll}
\hline
 & $t_{exp}$
 & 
 &
\\
Lens System
 & (sec)
 & $z_{\rm lens}$ 
 & References
\\
\hline
JVAS B0218+357  &     48720 & 0.685    & 1,2   \\
CLASS B0445+128 &      5228 & 0.557    & 3,4   \\
CLASS B0850+054 &      2296 & 0.59     & 4,5   \\
CLASS B1608+656 &      2528 & 0.630    & 6     \\
CLASS B2108+213 &      2304 & 0.365    & 7,8   \\
CFRS 03.1077    &      2296 & 0.938    & 9     \\
HE 0435-1223    &      1445 & 0.454    & 10,11 \\
HE 1113-0641    &      1317 & 0.75$^a$ & 12    \\
J0743+1553      &      2300 & 0.19     & 13    \\
J0816+5003      &      2440 & ...      & 14    \\
J1004+1229      &      2296 & 0.95     & 15    \\
RX J1131-1231   &      1980 & 0.295    & 16    \\
SDSS 0246-0825  &      2288 & 0.723    & 17,18 \\
SDSS 0903+5028  &      2444 & 0.388    & 19    \\
SDSS 0924+0219  &      1148 & 0.39     & 20,21 \\
SDSS 1004+4112  &      2025 & 0.68     & 22,23 \\
SDSS 1138+0314  &      2296 & 0.445    & 24    \\
SDSS 1155+6346  &      1788 & 0.176    & 25    \\
SDSS 1226-0006  &      2296 & 0.517    & 24    \\
WFI 2033-4723   &      2085 & 0.661    & 26,24 \\
\hline
\end{tabular}

\medskip
Redshifts marked with an $^a$ are photometric redshifts.  All other
redshifts are spectroscopic. \\
References:
 [1] \citet{0218_patnaik},
 [2] \citet{0218_browne}, 
 [3] \citet{0445_argo},
 [4] \citet{class_redshifts_mckean},
 [5] \citet{0850_biggs},
 [6] \citet{1608_myers},
 [7] \citet{2108_mckean},
 [8] \citet{2108_group},
 [9] \citet{cfrs03_crampton},
[10] \citet{0435_wisotzki},
[11] \citet{0435_morgan},
[12] \citet{1113_blackburne},
[13] \citet{0743_haarsma},
[14] \citet{0816_lehar},
[15] \citet{1004_lacy},
[16] \citet{1131_sluse},
[17] \citet{sdss0246_inada},
[18] \citet{qso_redshifts_eigenbrod_2},
[19] \citet{sdss0903_johnston},
[20] \citet{sdss0924_inada},
[21] \citet{sdss0924_eigenbrod},
[22] \citet{sdss1004_inada},
[23] \citet{SDSSJ1004_cluster},
[24] \citet{qso_redshifts_eigenbrod_1},
[25] \citet{sdss1155_pindor},
[26] \citet{wfi2033_morgan}.
\end{table}

\subsection{Control fields}

The first control sample consists of data obtained by the Cosmic
Evolution Survey team \citep[COSMOS;][]{cosmos_overview,cosmos_hst}.
The COSMOS data consist of a mosaic of approximately two square
degrees, with all of the images obtained through the F814W filter.
They were obtained as part of a 510-orbit HST Treasury proposal in
Cycles 12 and 13 (GO-9822; PI Scoville).  Each field has a total
exposure time of 2028~sec, comparable in depth to the lens fields. The
data have been fully reduced by the COSMOS team \citep{cosmos_reduc},
so it was not necessary to run {\em multidrizzle}.  Instead the
257 processed science and weight images from Cycle 12 -- comprising
$\sim$1 square degree -- were obtained from the COSMOS ACS website
hosted by the NASA/IPAC Infrared Science Archive\footnote{
{http://irsa.ipac.caltech.edu/data/COSMOS/images/acs\_v1.2/}}.

The second control sample consists of data obtained as part of a pure
parallel program to search random fields for emission line galaxies
(GO-9468; PI L.\ Yan).  This program included 28 pointings in the
F814W filter, of which we used the 20 that successfully passed all of
the criteria required for our image processing (hereafter referred to
as the ``pure parallel fields'').  These pure parallel fields cover a
total area of $\sim$0.06 square degrees.  Although the number of
pointings is much smaller than obtained with the COSMOS program, the
pure parallel fields have the advantage of not being contiguous on the
sky and, thus, provide an important check that the COSMOS data do not
have some overall bias in the number counts due to sample variance.
These data were processed by the a modified version of the pipeline
developed as part of the HST Archive Galaxy Gravitational Lens Survey
(Marshall et al., in prep) which is designed to produce final images
aligned sufficiently well to conduct weak lensing analyses.  The
modification to the pipeline for the pure parallel fields was simply
to change the output pixel scale from the HAGGLES standard 0\farcs03
pix$^{-1}$ to 0\farcs05 pix$^{-1}$, in order to match the COSMOS pixel
scale.

\begin{figure}
\includegraphics[width=\hsize]{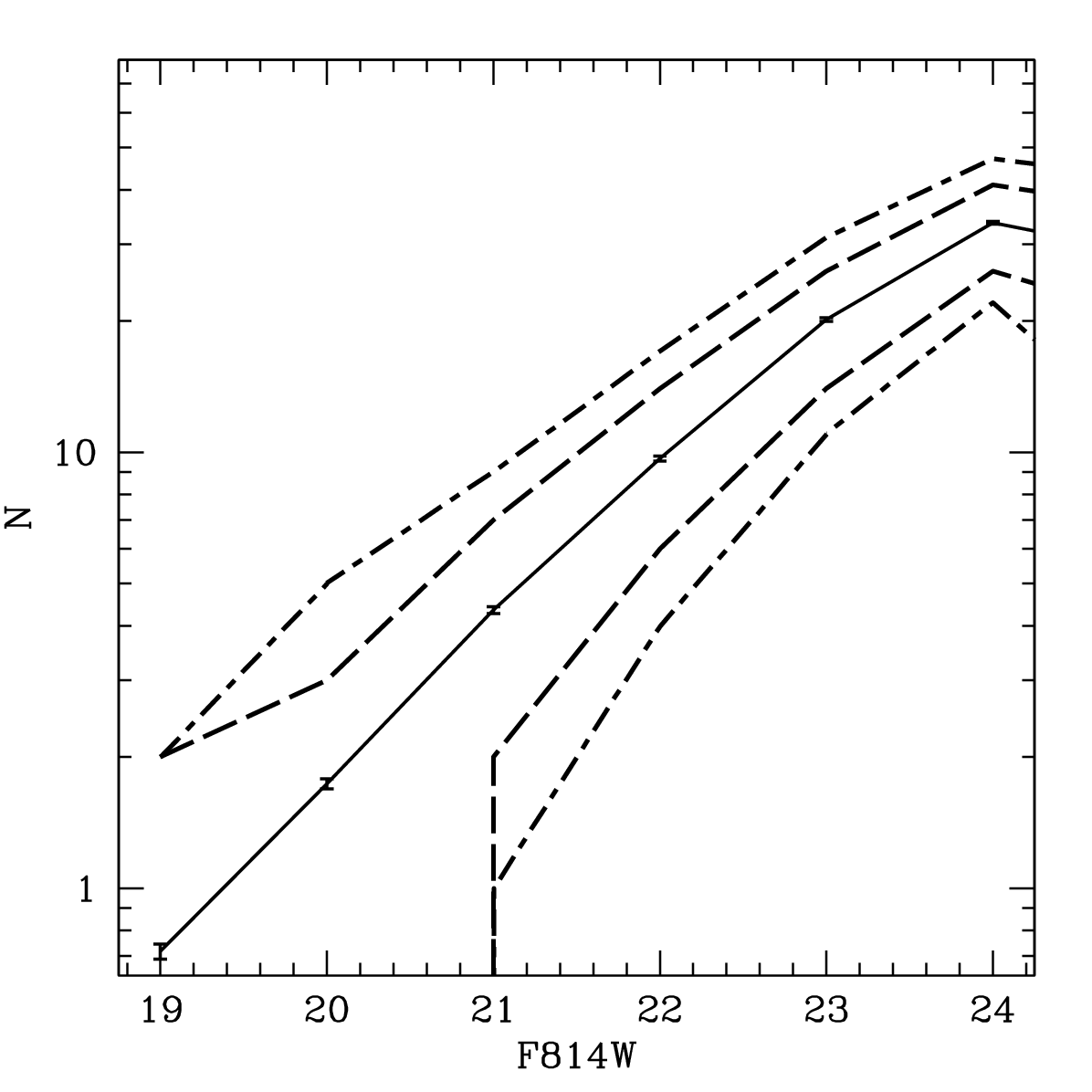}
\caption{
Number counts in the COSMOS fields, using apertures  of radius
45\arcsec.  There are four such apertures per COSMOS ACS field.
The thick solid line represents the mean number counts in each bin, while
the negligible error bars on the line show the formal error on the mean.  The
dashed and dot-dashed lines enclose 68\% and 90\% of the 
data, respectively. 
}\label{fig_cosmos_ns}
\end{figure}

\subsection{Object Detection and Flagging}

Object catalogs were obtained by running SExtractor \citep{sextractor}
on each ACS image.  For both the lens-field and control files, the
weight maps produced by {\em multidrizzle} were used to improve the
object detection.  The magnitudes used in this paper for all three
samples are Vega-based $F814W$ magnitudes measured within the ``AUTO''
aperture computed by SExtractor.  Hereafter, we will use $m$ to
designate these $F814W$ Vega magnitudes. We edited the initial
catalogs to reject stars and artifacts.  Because the {\em HST} point
spread function is simply the diffraction limit of the telescope
rather than depending on variable seeing, a simple star-galaxy
separation can be achieved by plotting the SExtractor full width at
half-maximum (FWHM) parameter versus object magnitude.  The stars
stand out as a narrow locus of objects all with approximately the same
FWHM, up until the point where they saturate (at $m \la 18 - 18.5$).
For brighter stars, the locus moves to larger values of the FWHM while
still remaining relatively narrow, making it easy to reject the stars.
However, this star-galaxy separation method does not catch false
detections due to stellar diffraction spikes and bleeding from
saturated regions.  Most of these objects can be still be flagged
automatically because they are often highly elongated.  Thus, to
select real galaxies from the catalogs, we rejected all sources with:
(1) FWHM$<0$\farcs13, (2) $m<18.5$, and (3) $(b/a)< 0.12$, where $a$
and $b$ are the semimajor and semiminor axes, respectively.

The COSMOS catalogs required additional flagging because the images
obtained from the COSMOS ACS science archive contain bands several
pixels wide along their left and right edges where cosmic rays are
not properly cleaned.  Simple spatial masks were sufficient to
eliminate the spurious sources associated with the cosmic rays.  The
COSMOS images were obtained at two fixed roll angles, separated by
180\degr\ \citep[e.g.,][]{cosmos_reduc}, so that two sets of masking
regions were required to flag the resulting catalogs.

\subsection{Definition of apertures}

We compute galaxy number counts in a set of apertures with radii of
45\arcsec/$(\sqrt{2})^i$, where $i = 0,1,2,3$.  These aperture sizes
are chosen to probe how localized any differences between the lens
fields and control fields may be.  The lens targets are centered on
one of the ACS chips, and the $i = 0$ aperture is roughly the largest
that can fit on the chip without extending over the chip gap.  For
apertures with $i>3$, the numbers of galaxies detected in the
apertures start to drop to unacceptably small numbers on the bright
end of the luminosity functions.  In the case of the lens fields, the
apertures are always centered on the lensing galaxy, while for the
control fields the apertures were laid down on regular grids to
maximize the number of independent apertures on each field while also
avoiding the chip edges and chip gap.  These grids consist of 4,
9, 16, and 36 apertures per pointing for aperture radii of 45\arcsec, 31\farcs8,
22\farcs5, and 15\farcs9, respectively.  Thus, for each choice
of aperture size there will always be 20 lens apertures, whereas
the number of control field apertures will depend on the aperture
size.  There will be ($257 \times n$) COSMOS and ($20 \times n$)
pure parallel apertures, where $n$ represents the aperture-dependent
number of grid points per pointing listed above.

Figure~\ref{fig_cosmos_ns} shows, as an illustration, the distribution
of galaxy number counts in the COSMOS fields inside apertures of
radius 45\arcsec, with lines marking the mean number counts
and the regions enclosing 68\% and 90\% of the data.
The formal errors on the mean are very small (nearly invisible on the
plot), but small number statistics significantly broaden the width of
the distribution at the bright end.  Furthermore, the 2000~s exposure
time for each COSMOS field leads to a turnover of the number counts
past $m = 24$.  Thus, in the following analysis we only consider
objects with $19\leq m \leq 24$.

Of course, having the lens-field apertures chosen to be centered on
the lens system introduces two sources of bias.  The first is that the
lens system itself, consisting of the lensing galaxy and the multiple
lensed images of the background source, contributes to the number
counts in the aperture.  The second bias is that lensing galaxies tend
to be massive early-type galaxies and, as such, can be expected to be
found in locally overdense environments
\citep[e.g.,][]{morphdens,zm98}.  We control the first bias by
flagging the lensing galaxy and all lensed images by hand in the input
catalogs.  Furthermore, we exclude any galaxies within $\theta =
$2\farcs5 of the lens systems to avoid any strong magnification biases
associated with these bright galaxies, although this is not expected
to be a large effect.  For typical galaxy-mass lenses, the Einstein
ring radii of $\theta_{\rm Ein} \sim$1\arcsec\ imply that the
magnifications of lensed images at $\theta = $2\farcs5 are $\sim$0.5
magnitude, and the magnification falls off as $\sim 1/\theta$.
Therefore, only a small solid angle around each lens produces
magnification at this level, suggesting that the number of faint
galaxies that are mistakenly placed into a brighter magnitude bin will
be minimal.  The correction for the second effect is discussed in
\S\ref{sec_xcorr_correction}.
There may be some additional bias in the number counts in the
lens fields due to clustering associated with the lensed background
object \citep[e.g.,][]{1608_serendip}, which for these systems is
almost always a massive galaxy hosting an active nucleus.  However, we
expect this bias in the counts to appear only in the fainter magnitude bins
and to be (1) mostly washed out by the large number of faint galaxies along
these lines of sight and/or (2) fainter than our $m = 24$ cutoff.


\begin{table}
\caption{KS Test Results}
\label{tab_ks}
\begin{tabular}{lllll}
\hline
              &         & \multicolumn{3}{c}{$\log_{10} P_{\rm KS}$} \\
$r$ (\arcsec) & $m$ & \multicolumn{1}{c}{Lens--COSMOS}
 & \multicolumn{1}{c}{PP--COSMOS} & \multicolumn{1}{c}{Lens--PP} \\
\hline
45.0 & 19 & -0.54       & -0.00       & -0.40 \\ 
     & 20 & {\bf -2.37} & -0.24       & -1.89 \\ 
     & 21 & -0.84       & -0.95       & -0.63 \\ 
     & 22 & -0.66       & {\bf -3.22} & -0.49 \\ 
     & 23 & -0.07       & {\bf -3.78} & -0.03 \\ 
     & 24 & -0.81       & -0.50       & -0.61 \\ 
31.8 & 19 & -0.52       & -0.00       & -0.45 \\ 
     & 20 & -1.93       & -0.12       & -1.73 \\ 
     & 21 & -0.94       & -1.69       & -0.83 \\ 
     & 22 & -0.11       & {\bf -8.77} & -0.08 \\ 
     & 23 & -0.03       & {\bf -5.60} & -0.02 \\ 
     & 24 & -0.60       & {\bf -2.04} & -0.52 \\ 
22.5 & 19 & -0.09       & -0.00       & -0.07 \\ 
     & 20 & -0.97       & -0.00       & -0.90 \\ 
     & 21 & -1.23       & -1.23       & -1.15 \\ 
     & 22 & -0.44       & {\bf -5.64} & -0.40 \\ 
     & 23 & -0.03       & -1.53       & -0.02 \\ 
     & 24 & -0.50       & -0.06       & -0.46 \\ 
15.9 & 19 & -0.03       & -0.00       & -0.03 \\ 
     & 20 & -0.23       & -0.00       & -0.22 \\ 
     & 21 & -0.90       & {\bf -2.83} & -0.87 \\ 
     & 22 & -0.60       & {\bf -8.18} & -0.58 \\ 
     & 23 & -0.30       & {\bf -6.43} & -0.29 \\ 
     & 24 & -0.13       & {\bf -5.28} & -0.13 \\ 
\hline
\end{tabular}

\medskip
Results from Kolmogorov-Smirnov tests comparing pairs of samples.
Values are the logarithms of the probabilities that the given pair of
samples could have produced the observed $D$ values by chance if they were 
drawn from the same distribution.  The ``PP'' designation refers to the pure
parallel sample.
Numbers in bold are those with probabilities lower than 0.01.
\end{table}

\section{Frequentist Analysis}

\begin{figure*}
\includegraphics[width=0.8\hsize]{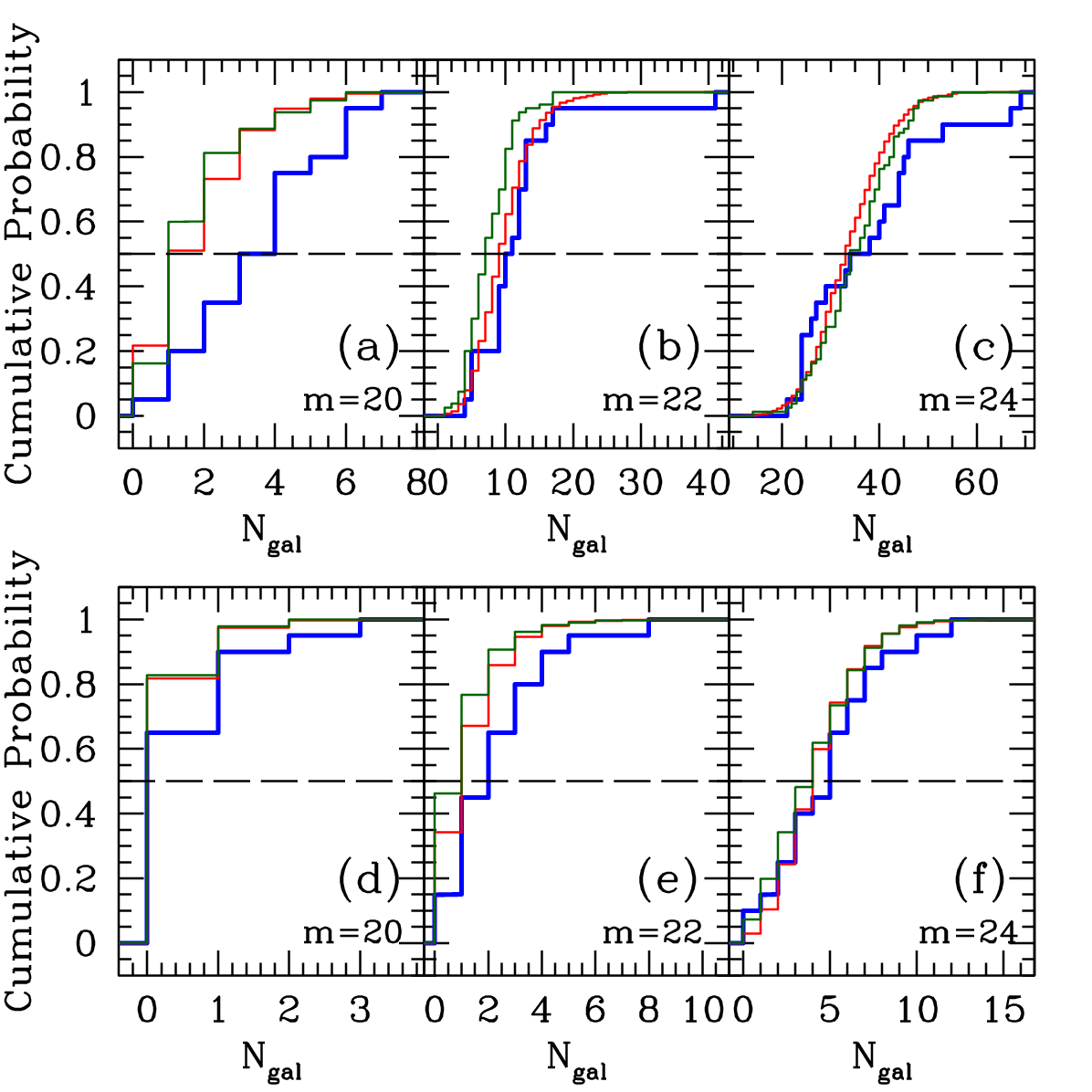}
\caption{ 
Cumulative distribution of number counts for six
representative aperture--magnitude pairs.  The magnitude bin used to construct each plot is
shown in the bottom right corner of the plot.  The horizontal dashed
line in each plot represents a cumulative probability of 0.5 and can
thus be used to find the medians of the distributions.  The thick blue curves
represent the lens sample.  The thin curves show the COSMOS sample (red)
and the pure parallel sample (green).  
{\bf (Top row)} 45\arcsec apertures.
{\bf (Bottom row)} 15\farcs9 apertures.
}\label{fig_cdf}
\end{figure*}

In order to do an initial comparison of the three samples, we use
Kolmogorov-Smirnov (KS) tests on three different pairs of samples:
lens vs.\ COSMOS, pure-parallel vs.\ COSMOS, and lens vs.\
pure-parallel.  These tests are conducted for each combination of
aperture size and magnitude.  Figure~\ref{fig_cdf} shows examples of
cumulative distributions for several combinations of aperture size and
magnitude bin.  These plots can be used to estimate the KS $D$ value
for representative pairs of samples, as well as the sample medians.
The results of the KS tests are given in Table~\ref{tab_ks}, and
reveal that for most of the aperture--magnitude pairs the lens fields
are consistent, at greater than the 10\% confidence level, with being
drawn from the same distribution as the control fields; only for the
comparison to the COSMOS sample in the $m = 20$ bin and the
45\arcsec\ aperture is the probability that the two samples are drawn
from the same distribution less than 1\%.  However, the significance
of any differences between the lens fields and the other samples is
low due to the small number of apertures in the lens fields.  With a
larger sample of lens targets, the differences in distributions may
become more significant.

Somewhat surprisingly, the control samples show evidence of
significant difference from each other, with 10 instances where the KS
test indicates that there is less than a 1\% chance that the
pure-parallel and COSMOS samples are drawn from the same parent
distribution.  Most of these low probabilities occur for the bins
where $m = 22$ or 23.  Furthermore, in nearly every case, it
appears that the COSMOS fields are overdense compared to the pure
parallel fields (e.g., Figure~\ref{fig_cdf}b).  While this is
an unexpected result, since both the COSMOS and the pure-parallel
fields were chosen to be ``fair'' representations of the Universe, it
is perfectly possible that the contiguous COSMOS area is not large
enough to escape being a biased line of sight through the Universe.
In fact, our results are consistent with analyses by the COSMOS team, which
find that, in the magnitude range $22 < i < 23$, the field chosen for
the COSMOS observations has higher clustering amplitudes than those
found in surveys of other fields \citep{mccracken}.  Also, the
COSMOS weak lensing maps \citep{cosmos_massey,cosmos_leauthaud}
show more structure than seen typical simulation fields \citep{faure_env}.
In this case the pure parallel sample, albeit small, appears to provide a
better indication of expected number counts in images of this depth.

\section{Bayesian Analysis}

To objectively compare the number counts in the lens and control
fields, and assess whether they are consistent, we
also conduct a Bayesian analysis of the number count distributions.

\subsection{Poisson Fluctuations}

\begin{figure*}
\centering
\includegraphics[width=0.7\hsize]{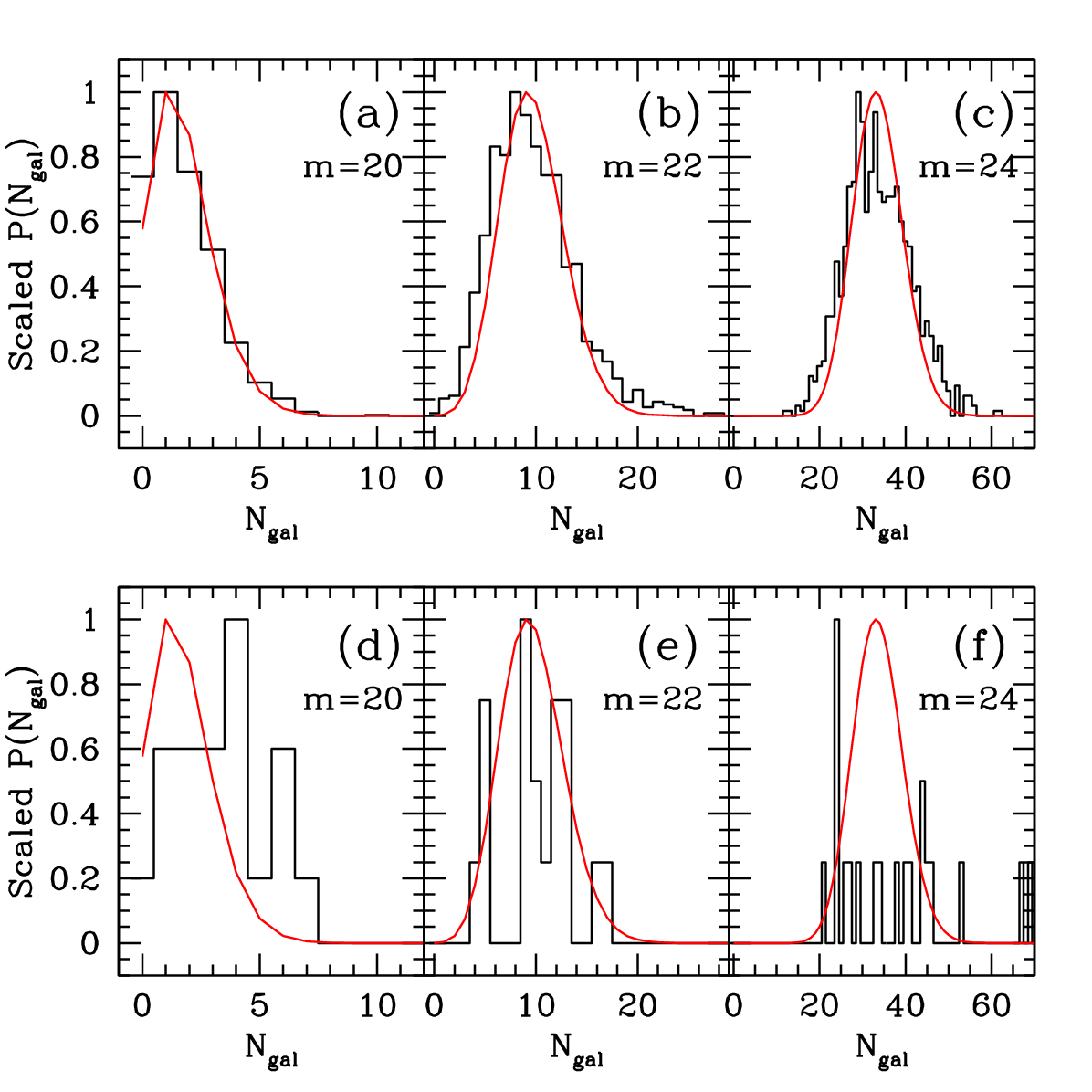}
\caption{
Histograms of number counts in the 45\arcsec aperture for two different 
samples: COSMOS (a--c)
and lens fields (d--f).  Shown are the magnitude bins $m=20,22,24$ from
right to left in each row.  In both rows, the histograms represent the
observed data while the solid curves represent a Poisson distribution
with the same mean {\em as the COSMOS data.}  Therefore, the Poisson
distributions for the lens fields may not have the same means as
the observed lens distributions.  Whereas for bright galaxies the
number-count variance can be explained by Poisson fluctuations, at
fainter magnitudes (i.e.\ higher number counts; see panels $b$ and $c$)
the effect of sample variance becomes
apparent. }\label{fig_obs_vs_poisson}
\end{figure*}

The first effect that needs to be considered in the Bayesian analysis
is that of counting statistics.  That is, given an underlying
expectation value, $\mu$, for the number of galaxies in a given
aperture and magnitude bin, what is the likelihood function for the
observed number counts, $N_i$, for each field $i$?  This is simply the
Poisson probability function
\begin{equation}
	P(N_i|\mu) = \frac{e^{-\mu} \mu^{N_i}}{N_i!},
\end{equation}
where $P(N_i|\mu)$ is already normalized. In the top row of
Figure~\ref{fig_obs_vs_poisson} we show representative plots that
include both the distributions of the observed COSMOS number counts
(histograms) and the corresponding Poisson distributions with the same
means (solid curves).  Figure~\ref{fig_obs_vs_poisson}a shows that the
Poisson description works well for distributions with small means, in
this case a bright-magnitude bin for the 45\arcsec\ aperture.  In
contrast, as $\mu$ becomes large (e.g.,
Figure~\ref{fig_obs_vs_poisson}c), the observed distribution becomes
wider than the predicted one, suggesting that different apertures have
different underlying density fields, i.e., the same value of $\mu$
cannot be used for all apertures of a given size.  Clearly the
analysis requires another term.

\subsection{Sample (``Cosmic'') Variance}

The additional term is necessary because the presence of large-scale
structure produces field-to-field variations, i.e.  sample variance,
in the expectation value of the underlying density field.  To model
this large-scale structure term, commonly described as ``cosmic variance'',
we assume that the field-to-field variations, for a fixed aperture,
within the lens and control samples can, in each case, be approximated
as a Gaussian random field \citep{bardeen} for $\mu>0$, i.e.
\begin{equation}
   P(\mu| \mu_0, \sigma_0) \propto 
       \exp\left(-\frac{(\mu-\mu_0)^2}{2 \sigma_0^2}\right) 
\end{equation}
and $P(\mu| \mu_0, \sigma_0)=0$ for $\mu<=0$. The integral over the
probability function is properly normalized to unity. The values of $\mu_0$ and
$\sigma_0$ for a given aperture size are held fixed {\em within} each ensemble
of fields (i.e., lens, COSMOS, or pure parallel) but can vary {\em between}
ensembles.

\begin{figure*}
\includegraphics[width=0.8\hsize]{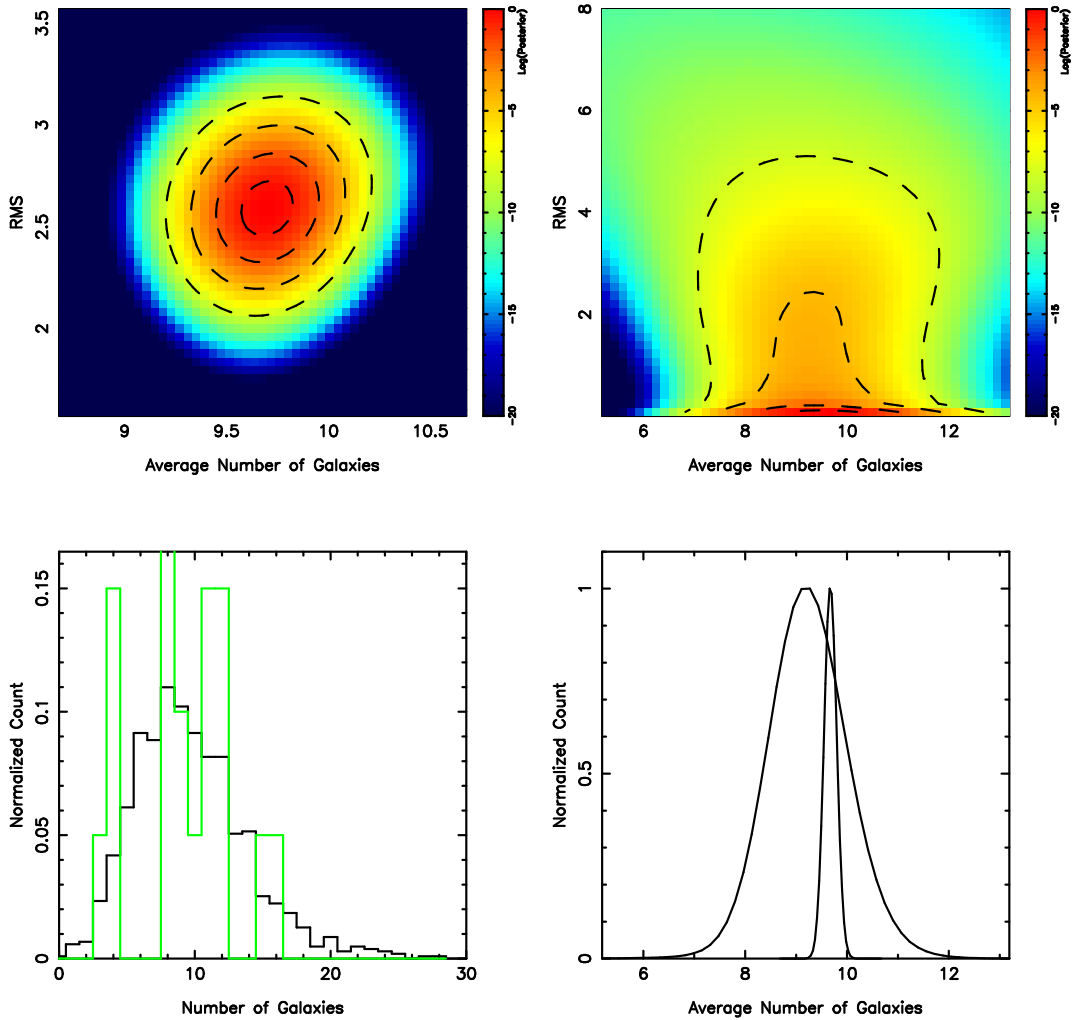}
\caption{
Bayesian inference of the mean global galaxy number density, $\mu_0$,
for the lens and control fields.  Shown is the result for the largest
(45\arcsec) aperture and $m=22$ magnitude bin. Shown in the lower-left
panel are the (normalized) number counts of lens (green) and control
fields (black; COSMOS in this case) as function of the number of
galaxies in those fields. The upper two panels show the posterior
probability density of $(\mu_0, \sigma_0)$ for the lens (right) and
control (left) fields, as determined from equation 4. The lower-right
panel shows the marginalized probability function of $\mu_0$ for the
lens (broad function) and control (narrow function) fields.
}
\label{fig_bayesian_example}
\end{figure*}

Using Bayesian theory to combine the two effects gives
\begin{equation}
	P(N_i|\mu_0, \sigma_0) = \int P(N_i|\mu) P(\mu| \mu_0, \sigma_0) d\mu,
\end{equation}
and, combining the different fields within a given sample into a
single data set $\{N_i\}$, with
\begin{equation}	
	P(\mu_0, \sigma_0 | \{N_i\}) = \frac{P(\mu_0, \sigma_0) \prod_i P(N_i|\mu_0, \sigma_0)  }{P(\{N_i\})}.
\end{equation}
We assume a flat prior on $\mu_0$, because it must be invariant under
shifts, and a flat prior on $\log(\sigma_0)$, because it must be
invariant under multiplication \citep[e.g.,][]{gregory}.
Finally, we marginalize over $\sigma_0$, which to us is a nuisance
parameter, and get
\begin{equation}
	P(\mu_0 | \{N_i\}) = \int P(\mu_0, \sigma_0 | \{N_i\}) d\sigma_0.
\end{equation}
To obtain the median and 68\% confidence contour, we construct a
marginalized probability distribution from $P(\mu_0 | \{N_i\})$ and
find the $\mu_0$ values corresponding to cumulative probabilities of
0.16, 0.5 (median), and 0.84.  In Fig.~\ref{fig_bayesian_example}, we
illustrate the above process by an example for the 45\arcsec\ aperture
and for $m = 22$.

\begin{table*}
\bigskip
\centering
\caption{Estimated value of the mean underlying number counts $\mu_0$ 
in the lens, COSMOS, and pure-parallel fields. \label{tab_mu0}}

\begin{tabular}{ccrrrrr}
\hline
Radius
 & m
 & \multicolumn{1}{c}{$\mu_0$}
 & \multicolumn{1}{c}{$\mu_0$}
 & \multicolumn{1}{c}{$\mu_0$}
 & Correlation Function
 & Corrected \\
(Arcsec.)
 & (mag.)
 & \multicolumn{1}{c}{(COSMOS)}
 & \multicolumn{1}{c}{(Pure-parallel)}
 & \multicolumn{1}{c}{(Lens)}
 & \multicolumn{1}{c}{Correction ($\Delta \mu_0$)}
 & \multicolumn{1}{c}{$\mu_0$ (Lens)}\\
\hline
45.0 &19 &  $0.67_{-0.07}^{+0.04}$ &   $0.5_{-0.3}^{+0.2}$    &  $0.5_{-0.2}^{+0.3}$ & $0.15 \pm 0.08$ & $ 0.4_{-0.2}^{+0.3}$ \\
     &20 &  $1.70 \pm 0.05$        &   $1.6_{-0.1}^{+0.2}$    &  $2.5 \pm 0.5$       & $0.22 \pm 0.11$ & $ 2.3 \pm 0.5      $ \\
     &21 &  $4.32 \pm 0.08$        &   $3.5 \pm 0.2$          &  $3.8 \pm 0.5$       & $0.33 \pm 0.17$ & $ 3.5 \pm 0.5      $ \\
     &22 &  $9.7  \pm 0.1$         &   $7.5 \pm 0.4$          &  $9.2_{-0.7}^{+0.8}$ & $0.44 \pm 0.22$ & $ 8.7_{-0.7}^{+0.8}$ \\
     &23 & $20.1  \pm 0.2$         &  $17.6 \pm 0.6$          & $19.4 \pm 1.7$       & $0.56 \pm 0.28$ & $18.9 \pm 1.7$       \\
     &24 & $33.5  \pm 0.2$         &  $34.9_{-0.5}^{+0.6}$    & $36.4 \pm 3.0$       & $0.56 \pm 0.28$ & $35.8 \pm 3.0$       \\ \hline
31.8 &19 &  $0.32_{-0.11}^{+0.02}$ &   $0.2 \pm 0.1$          &  $0.4_{-0.2}^{+0.3}$ & $0.10 \pm 0.05$ & $ 0.3_{-0.2}^{+0.3}$ \\
     &20 &  $0.85_{-0.04}^{+0.03}$ &   $0.7_{-0.2}^{+0.1}$    &  $1.0  \pm 0.3$      & $0.14 \pm 0.07$ & $ 0.9 \pm 0.3      $ \\
     &21 &  $2.14 \pm 0.03$        &   $1.7 \pm 0.1$          &  $2.0_{-1.3}^{+1.1}$ & $0.22 \pm 0.11$ & $ 1.8_{-1.3}^{+1.1}$ \\
     &22 &  $4.90 \pm 0.06$        &   $3.5 \pm 0.3$          &  $2.5_{-1.7}^{+2.2}$ & $0.29 \pm 0.15$ & $ 2.2_{-1.7}^{+2.2}$ \\
     &23 & $10.10 \pm 0.08$        &   $8.6 \pm 0.3$          &  $9.6_{-2.0}^{+1.6}$ & $0.37 \pm 0.18$ & $ 9.2_{-2.0}^{+1.6}$ \\
     &24 & $16.64 \pm 0.1$         &  $15.8 \pm0.4$           & $17.3_{-2.0}^{+1.9}$ & $0.37 \pm 0.19$ & $16.9_{-2.0}^{+1.9}$ \\ \hline
22.5 &19 &  $0.04_{-0.02}^{+0.10}$ &   $0.11_{-0.08}^{+0.06}$ &  $0.4_{-0.3}^{+0.4}$ & $0.06 \pm 0.03$ & $ 0.4_{-0.3}^{+0.4}$ \\
     &20 &  $0.42_{-0.06}^{+0.01}$ &   $0.38 \pm0.04$         &  $0.3_{-0.2}^{+0.3}$ & $0.09 \pm 0.05$ & $ 0.2_{-0.2}^{+0.3}$ \\
     &21 &  $1.03 \pm 0.02$        &   $0.7 \pm 0.1$          &  $1.7_{-0.6}^{+0.5}$ & $0.14 \pm 0.07$ & $ 1.5_{-0.6}^{+0.5}$ \\
     &22 &  $2.42 \pm 0.03$        &   $1.8 \pm 0.1$          &  $1.8_{-1.2}^{+1.4}$ & $0.19 \pm 0.10$ & $ 1.6_{-1.2}^{+1.4}$ \\
     &23 &  $5.04 \pm 0.04$        &   $4.5 \pm 0.2$          &  $3.8_{-1.9}^{+1.6}$ & $0.24 \pm 0.12$ & $ 3.6_{-1.9}^{+1.6}$ \\
     &24 &  $8.37 \pm 0.05$        &   $8.4 \pm 0.2$          &  $8.6_{-1.1}^{+1.0}$ & $0.24 \pm 0.12$ & $ 8.4_{-1.1}^{+1.0}$ \\ \hline
15.9 &19 &  $0.08 \pm 0.01$        &   $0.07_{-0.04}^{+0.02}$ &  $0.7_{-0.4}^{+0.6}$ & $0.04 \pm 0.02$ & $ 0.6_{-0.4}^{+0.6}$ \\
     &20 &  $0.10 \pm 0.01$        &   $0.18_{-0.04}^{+0.02}$ &  $0.5_{-0.3}^{+0.4}$ & $0.06 \pm 0.03$ & $ 0.4_{-0.3}^{+0.4}$ \\
     &21 &  $0.48 \pm 0.01$        &   $0.12_{-0.07}^{+0.20}$ &  $0.9 \pm 0.4$       & $0.09 \pm 0.05$ & $ 0.8 \pm 0.4$      \\
     &22 &  $1.14 \pm 0.02$        &   $0.4 \pm 0.1$          &  $1.2_{-0.7}^{+0.6}$ & $0.13 \pm 0.06$ & $ 1.0_{-0.7}^{+0.6}$ \\
     &23 &  $2.51 \pm 0.02$        &   $1.9 \pm 0.1$          &  $2.3_{-1.1}^{+0.6}$ & $0.16 \pm 0.08$ & $ 2.1_{-1.1}^{+0.6}$ \\
     &24 &  $4.18 \pm 0.02$        &   $3.8 \pm 0.1$          &  $3.9_{-1.1}^{+0.7}$ & $0.16 \pm 0.08$ & $ 3.8_{-1.1}^{+0.7}$ \\
\hline
\end{tabular}
\end{table*}

\subsection{Results of Bayesian Analysis}

\begin{figure*}
\centering
\includegraphics[width=\hsize]{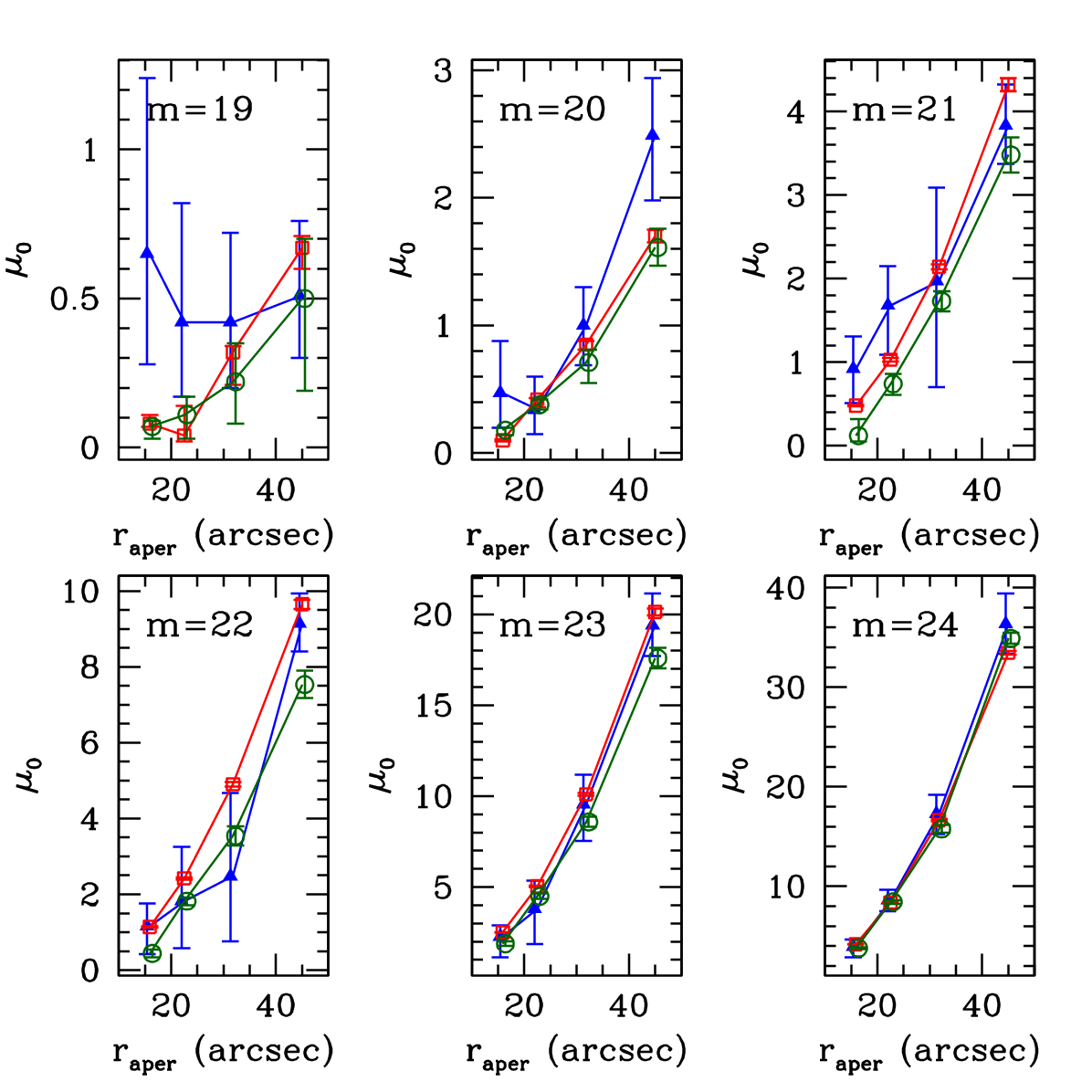}
\caption{Plots of $\mu_0$ vs. aperture size for each magnitude bin.
In each plot, the blue triangles represent the lens sample, the red
open squares represent the COSMOS sample, and the green open
circles represent the pure-parallel sample.  The value of $\mu_0$
is determined for each aperture independently.  Thus, for small
numbers of galaxies, such as in the $m = 19$ bin, it is possible
to have a larger fitted value of $\mu_0$ in a smaller aperture
than in the next larger aperture.  Note that the COSMOS
and pure-parallel points are formally inconsistent in several of
the plots (e.g., m=21, 22, and 23).}
\label{fig_aperture_counts}
\end{figure*}

We calculate $\mu_0$ 
for the lens and control fields as a function of the
aperture size and magnitude.  The results are shown in Figure
\ref{fig_aperture_counts} and listed in Table~\ref{tab_mu0}. 

Figure~\ref{fig_aperture_counts} shows significant differences between
the values of $\mu_0$ obtained for the two control samples.
Especially in the magnitude bins corresponding to $21 \leq m \leq
23$, the COSMOS galaxy densities are systematically higher than those
seen in the pure parallel fields.  This result is similar to that
obtained from the KS analysis (Table~\ref{tab_ks} and
Figure~\ref{fig_cdf}) and also with the analysis by the COSMOS team,
which finds that the amount of structure in the COSMOS field is at the
high end of the range of variations produced by sample variance
\citep{mccracken}.

\begin{figure*}
\centering
\includegraphics[width=0.7\hsize]{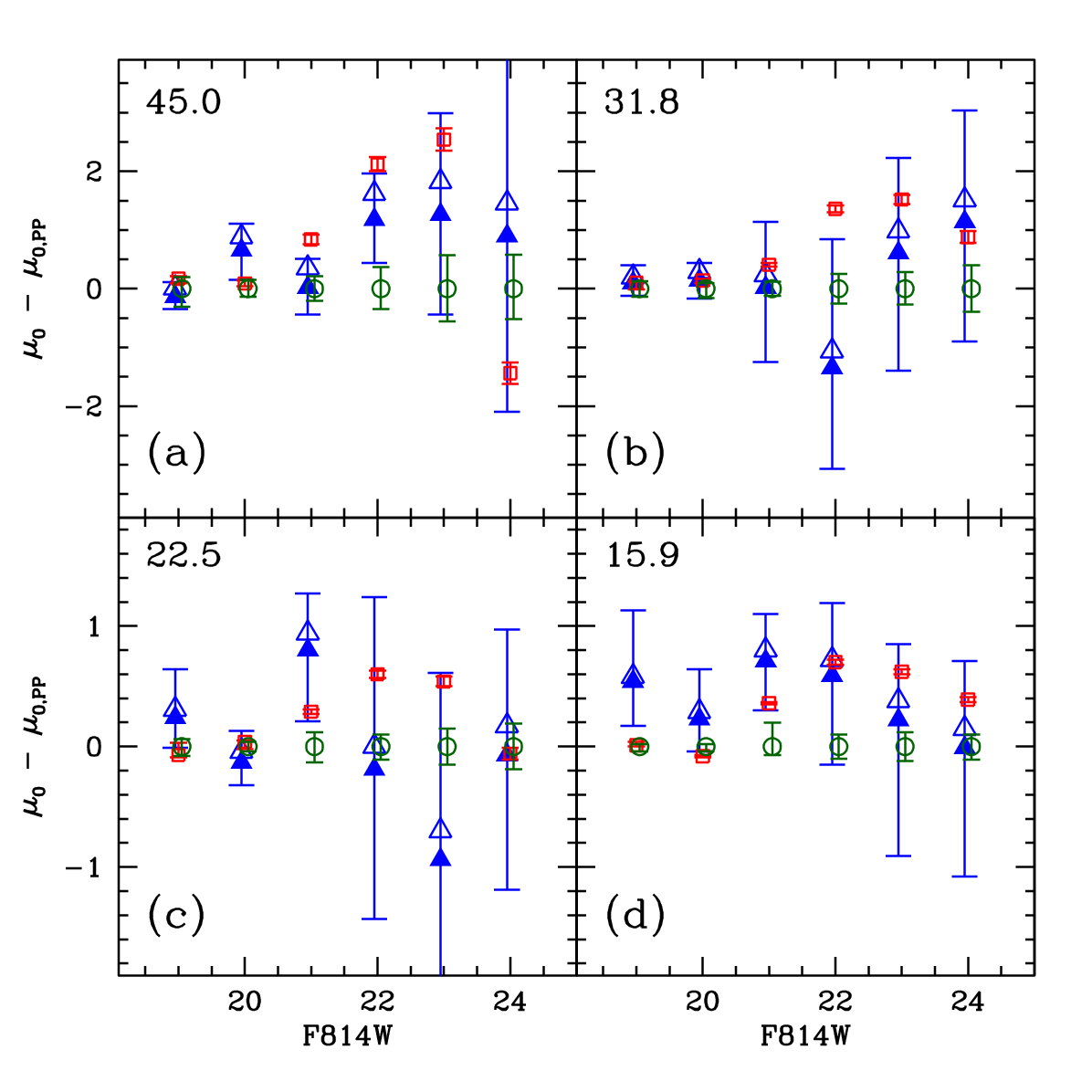}
\caption{
Results of Bayesian analysis, showing the {\em difference} between the
values of $\mu_0$ calculated for lens (blue triangles) and COSMOS fields
(red squares) fields and those obtained for the the pure parallel fields (green
circles) .  The open blue triangles represent the lens-field
values {\em before} correcting for clustering of massive galaxies, while
the solid blue triangles represent the lens-field values {\em after} the
correction has been applied (see \S\ref{sec_xcorr_correction}).  Note the
highly significant displacement between the pure parallel fields and the
COSMOS fields, particularly for $21 \leq m \leq 23$.
}\vspace{1cm}
\label{fig_delta_mu0}
\end{figure*}

It is also instructive to plot the results in terms of the offsets
in $\mu_0$ with respect to one of the samples.  For this exercise, the
fiducial sample is set to the pure-parallel sample because the COSMOS
field appears to be biased high.  Figure~\ref{fig_delta_mu0} shows the
resulting offsets in $\mu_0$.  Two trends are seen in the plot: (1)
For all apertures, the COSMOS-field values of $\mu_0$ are higher than
the pure-parallel values in the range $21 \leq m \leq 23$, often at
high significance, and (2) the lens fields often appear overdense
compared to the pure-parallel fields, but the uncertainties on the
lens values are so large that very few of the offsets are
significantly different from zero.  This result is consistent with our
frequentist analysis of the data (\S3).  Clearly, a much larger lens
sample is needed in order to evaluate whether the differences in
number counts are real or are mostly due to statistical fluctuations
in a small data set. Only for the smallest aperture (15\farcs9) do we
see a significant trend for the lens fields, with the three bins at
$m<22$ all being $\geq 1\sigma$ higher than the pure parallel values
(before the correlation function correction [\S4.4]).
To quantify the differences between the fields, we integrate the
expectation values in Table 3 over magnitude ($19 \leq m \leq 24$) to
obtain the typical number of galaxies ($N_{\rm int}$) in each sample.
Taking $\Delta N_{\rm int} \equiv N_{\rm int,lens} - N_{\rm int,pp}$,
we find that $\Delta N_{\rm int} = 6.1\pm 3.7$, $2.2\pm 3.8$, $0.7 \pm
2.8$ and $ 2.9 \pm 1.9$ galaxies for the 45\arcsec, 31\farcs8,
22\farcs5 and 15\farcs9 apertures, respectively.  In each case, the
error is dominated by the error in $N_{\rm int,lens}$.

Overall, we can conclude that over a wide range of apertures ($\leq
45$\arcsec), the difference in the number of galaxies between lens and
control fields is less than $\sim$6 galaxies. Although this can be
fractionally large, it clearly shows that on average only a few
galaxies determine the difference between lens fields and non-lens
fields in typical observations. For the smallest aperture of
15\farcs9, the difference is typically $\le1$ galaxy at the 68\%
confidence level in any given magnitude bin. However, since these
galaxies are, by definition, at small projected radii from the lens,
they have most influence on the lensing potential.  It may be
appropriate to explicitly include these galaxies in the lens model,
based on an evaluation of their estimated mass and projected distance
to the lens.

\subsection{Correlation-function Corrections \label{sec_xcorr_correction}}

In the previous subsection we saw that over the entire range of
aperture sizes and magnitude limits, the average difference in the
number of galaxies seen in the lens and control fields is typically
$\la 6$ for aperture sizes of 45\arcsec, integrated over $19 \leq m \leq 24$
(e.g., Fig.~\ref{fig_delta_mu0}).  
%
The slight excess seen in the lens fields is not unexpected because the
massive lens galaxies typically reside in {\em locally} overdense
regions and, in principle, one would expect more galaxies in their
neighborhood.

To quantify the effect of local overdensities, we use a two-point
correlation function.  We assume the functional form
$\omega(\theta)=A_{\omega} (\theta/1')^{-\delta}$ with $\delta=0.8$
from \citet{mccracken}.  Although both our analysis and that conducted
by \citet{mccracken} use F814W magnitudes, they use AB magnitudes
while we use Vega magnitudes.  Therefore, we subtract 0.42 from their
F814W AB magnitudes to obtain Vega magnitudes\footnote{ The ACS
zeropoints for the Vega and AB systems can be obtained at
http://www.stsci.edu/hst/acs/analysis/zeropoints.}. It appears
that, given the errors, a linear correlation between
$\log(A_{\omega})$ and F814W magnitude is a fairly good description of
the normalization constant with $\log(A_{\omega}) = -1.0\pm 0.25$ at
$m=19$ and $\log(A_{\omega}) = -2.1\pm 0.1$ at $m=24$, covering the
entire observed magnitude range.  We now have a functional form that
allows us to determine the expected overdensities in the lens fields
by integrating over $\omega(\theta)$ from 2.5 arcsec (our inner
cutoff) to the aperture radius and multiplying this by the average
density in the field.  Although the fractional uncertainties on the
normalization constants are large for the bright magnitudes, the
numbers of galaxies in these bins is small.  Thus, the uncertainties
on the predicted number of additional galaxies will be $<1$ galaxy.

In addition to the uncertainties in $\log(A_{\omega})$, there is an
uncertainty on the estimated number of excess galaxies that comes from
the value used for the average density in the field.  We have assumed
that the average galaxy densities over the full COSMOS and
pure-parallel field areas provide a reasonable approximation since
they are either derived from a very large field or from randomly
pointed fields, respectively. Even if they themselves are slightly
over- or underdense on these scales, we do not expect the shape of the
correlation function to be significantly altered.  Thus, the effect
of, say, the overdensity seen in the COSMOS field (e.g,
Fig.~\ref{fig_delta_mu0}) is to produce a slight overestimate on the
correction ($\Delta \mu_0$).  In response to this potential
overcorrection we broaden the errors on $\Delta \mu_0$.  Because this
is a correction on a correction, however, its effect is only
second-order.
Given the errors on the normalization, we conservatively estimate
(assuming a variation up and down by 1--$\sigma$ in all magnitude
bins) an upper limit for the error on $\Delta \mu_0$ of $\sim$50\%.
We use this upper limit for the uncertainties on all subsequent
estimates of the total number of additional galaxies.

\begin{figure*}
\centering
\includegraphics[width=0.7\hsize]{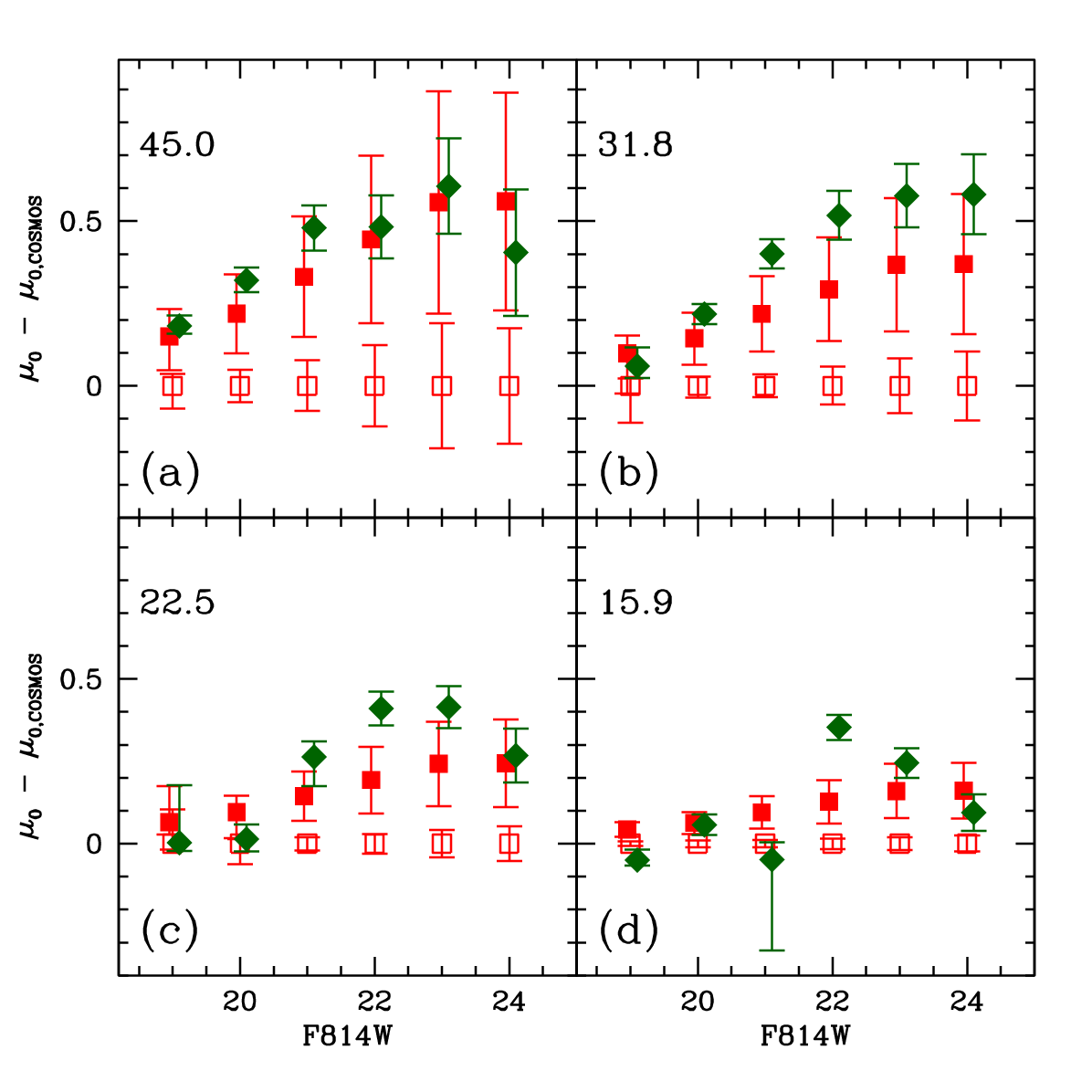}
\caption{
Comparison of number counts between the COSMOS control sample
(open red squares) and apertures centered on bright galaxies in COSMOS
(green diamonds).  The filled red squares represent the sum of the control
sample number counts and the two-point correlation correction
factors.
}\vspace{1cm}
\label{fig_bgal}
\end{figure*}

To test the validity of the two-point correlation correction, we
compared two sets of number counts derived from COSMOS.  The first set
is the one that we have used as the the first control sample, with
number counts computed in grids of apertures placed on each COSMOS
field.  This set should be considered to represent ``random'' lines of
sight through the COSMOS fields, since there is nothing special about
the location of the aperture centers.  In contrast, the apertures in
the second set are each centered on a bright ($m < 20.5$) galaxy found
in the COSMOS area.  The bright-galaxy sample contains 1801 apertures.
In Figure~\ref{fig_bgal}, we plot (1) the COSMOS ``random'' number
counts, (2) the bright-galaxy number counts, and (3) the sum of the
``random'' number counts and the two-point correlation correction.
The corrected random number counts are in excellent agreement with the
number counts in apertures centered on bright galaxies.

We can integrate the correlation-function corrections over magnitude
($19 \leq m \leq 24$) to estimate the total {\em excess} number of
galaxies, $\Delta \mu_{0,{\rm int}}$, that are expected in each of the
aperture sizes.  This calculation gives $\Delta \mu_{0,{\rm int}}$ =
2.0$\pm$1.0, 1.3$\pm$0.7, 0.9$\pm$0.5, and 0.6$\pm$0.3 galaxies, in
the 45\arcsec, 31\farcs8, 22\farcs5 and 15\farcs9 apertures,
respectively.  We subtract these values from the observed lens-field
excesses to correct for clustering, and find integrated differences of
$\Delta N_{\rm int,corrected} = 4.1\pm 3.8$, $0.9\pm 3.9$, $-0.2 \pm
2.8$ and $ 2.3 \pm 1.9$, respectively. In other words, after
correcting for galaxy clustering, the lens and pure-parallel (or
COSMOS) fields differ at less than the 2-$\sigma$ level.

%
%

%

\section{Interpretation}

Drawing firm conclusions from our Bayesian analysis is hampered by the
small number of lens fields in our sample.  There can be two
interpretations of our results: (1) that any line-of-sight
overdensities in the lens fields are insignificant once the local
overdensities around the lens galaxies are accounted for, and will
remain insignificant even with a large increase in the size of the
lens-field sample, or (2) the overdensities we see in $N_{\rm
int,corrected}$ for, e.g., the 45\arcsec\ and 15\farcs9 apertures will
become significant once more lens fields can be included in the
analysis and the errors on the estimates of $\mu_{0,{\rm lens}}$
shrink.  Distinguishing between the two will clearly need a larger
lens sample.  However, it should be noted that even if a future larger
lens sample indicates that the differences between the lens and
control fields are significant, the excess number of galaxies in the
lens fields is only, on average, $\Delta \mu_{0,{\rm int}} \la 4$ for
average integrated number counts of $N_{\rm int} \sim 70$; i.e., $\sim$6\%.

\subsection{Comments on Individual Lens Systems}

\begin{figure}
\centering
\includegraphics[width=0.9\hsize]{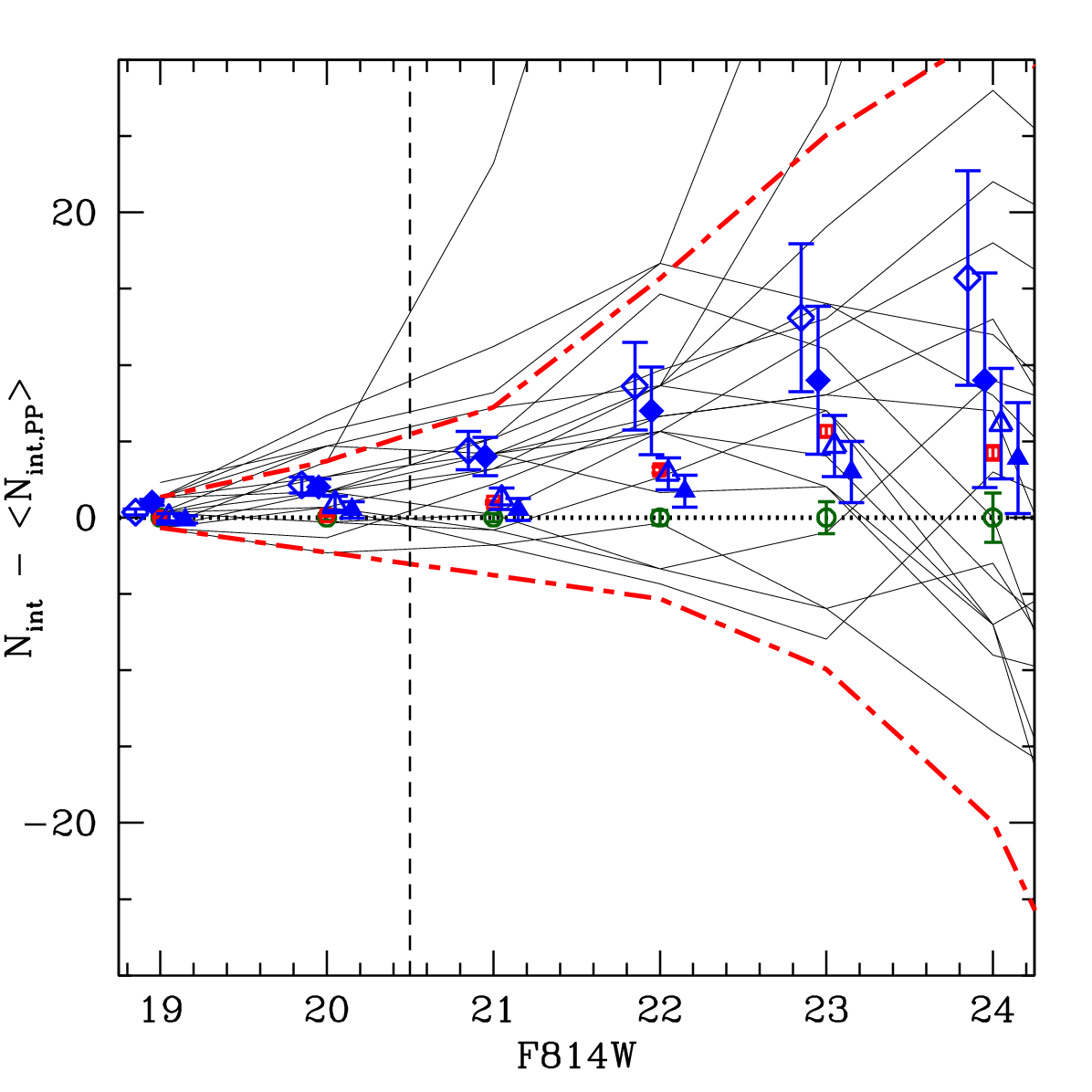}
\caption{
Offsets between the cumulative number counts calculated in individual
lens fields (light lines) and the pure-parallel mean cumulative number
counts (green circles) in apertures of radius 45\arcsec.  The blue points
represent four ways of representing the lens-field distribution: the
mean (open diamonds), the median (filled diamonds), $\mu_{0,{\rm int}}$ from the
Bayesian analysis (open triangles), and the corrected $\mu_{0,{\rm int}}$
(filled triangles).  The blue points have been offset slightly in the
horizontal direction for clarity.
The COSMOS distribution is also shown by its mean (red squares),
and 90\% range (red dot-dashed lines).
The light dashed vertical line is placed
solely to identify the three most overdense lens fields.  Starting at
the top of the figure and going down, the line encounters, in order,
the curves for SDSS J1004+4112, B1608+656, and B2108+213, all of which
fall outside the 90\% range of COSMOS field number counts.}
\label{fig_lens_intdiff}
\end{figure}

\begin{figure}
\centering
\includegraphics[width=0.9\hsize]{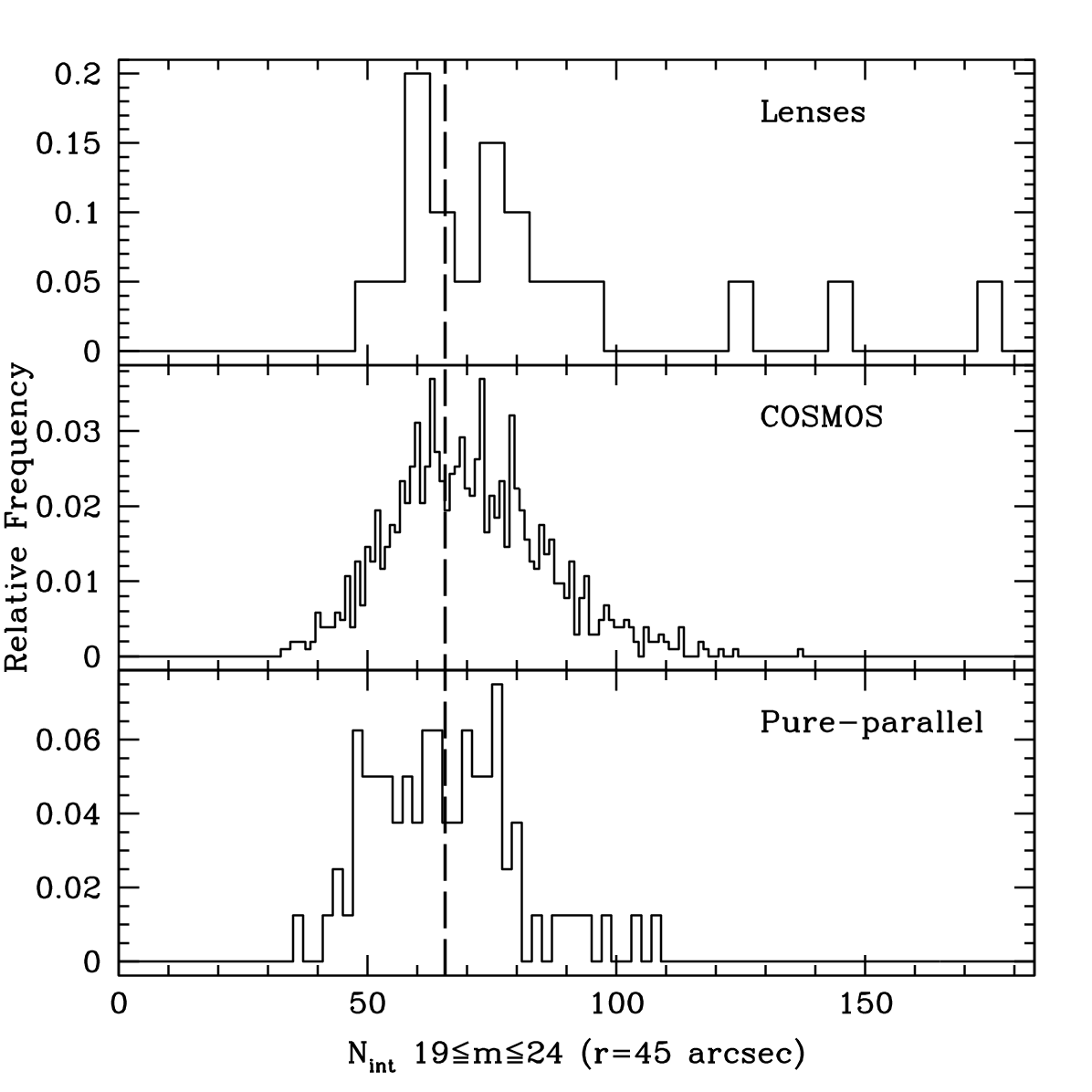}
\caption{Distribution of the integrated
number counts for all galaxies with $m \leq 24$ inside 45\arcsec\ apertures,
for each of the three samples.  The vertical dashed line represents the
mean of the pure-parallel sample.
}
\label{fig_lens_vs_cosmos}
\end{figure}

Of course, the average values that have been determined in the
preceding sections can hide a large variation from lens system to lens
system.  Fig.~\ref{fig_lens_intdiff} shows the difference between the
cumulative number counts of the lens sample and the pure-parallel mean
cumulative counts ($\Delta N_{\rm int}$) for the 45\arcsec\ aperture,
.  For the bright galaxies ($m \leq 22$), the lens fields with the
largest overdensities are, in order starting with the most overdense,
SDSS~J1004+4112, B1608+656, and B2108+213.  If considering all
galaxies with $m \leq 24$, the three most overdense fields are
SDSS~J1004+4112, B1608+656, and B0218+357 (the three highest in
Fig.~\ref{fig_lens_vs_cosmos}). All of these fields are outside the
region enclosing 90\% of the COSMOS data.  The overdensities for
SDSS~J1004+4112 and B2108+213 are not particularly surprising because
both of these systems are known to be associated with a cluster or
rich group \citep{SDSSJ1004_cluster,2108_group}.  However, neither
B1608+656 nor B0218+357 appear to be physically associated with such
massive concentrations of galaxies.  Spectroscopic observations of the
B1608+656 field have revealed multiple group-sized associations along
the line of sight \citep{1608groupdisc}, but the B0218+357 field will
have to be examined more closely in future analyses.
Fig.~\ref{fig_lens_intdiff} also reveals fields that are underdense
with respect to the mean.  The most underdense in bright galaxies are
SDSS~1226$-$0006, J0816+5003, and B0850+054.

The distributions of integrated lens number counts shown in
Figures~\ref{fig_lens_intdiff} and \ref{fig_lens_vs_cosmos} reveal
some interesting points.  One is that there are significant more high
outliers in the lens distribution than would be expected given the
control-field distributions.  This effect is almost certainly due in
part to small number statistics, and the highest point in the
integrated number count distribution in Fig.~\ref{fig_lens_vs_cosmos}
is due to an obvious cluster lens.  However, the next two highest
points in that distribution are due to galaxy-scale lenses that were
selected in a radio survey targeting the background objects
\citep{class_2}.  Because these lens systems were discovered in a
source-selected survey, they should not be biased toward high number
counts in the way that surveys targeting the likely lensing galaxies
(i.e., massive ellipticals) could be.  The second point, somewhat
related, is that our Bayesian analysis has produced clearly different
estimates of the ``typical'' number counts than other measures such as
the sample mean or median for the lens sample
(Fig.~\ref{fig_lens_intdiff}).  This is in contrast to the behaviour in
the control fields, where the sample means are excellent matches to
the calculated values of $N_{\rm int}$.  Clearly the mean of the lens
sample is pulled high by the outliers, but the median should be a more
robust estimator.  A significantly larger lens sample is needed to
assess whether these behaviours are due to small-number statistics or
are indicative of a slight bias in the lens-field counts.

\subsection{Implications for $H_0$}

Because B0218+357 \citep{biggs_0218delay}, SDSS~J1004+4112
\citep{J1004_delay1,J1004_delay2}, and B1608+656
\citep{1608_delay1,1608_delay2} are systems for which time delays have
been measured, the galaxy overdensities 
must be included in any analysis to determine $H_0$ from these
systems.  The over- and underdensities in mass along the line of
sight to a lens system, quantified by the external convergence
($\kappa_{\rm ext}$), bias the determination of $H_0$ from that
system if not taken properly into account.  The correct value of $H_0$
from a given field is $H_0 = (1 - \kappa_{\rm ext}) H_{0,{\rm
uniform}}$, where $H_{0,{\rm uniform}}$ is the value of $H_0$ obtained
without taking into account the external convergence.  In theory, it
should be possible to use our number-count approach to obtain an
estimate of $\kappa_{\rm ext}$ for a given lens.  The problem, of
course, is how to convert the observed galaxy numbers into an accurate
mass measurement.

One approach to estimate $\kappa_{\rm ext}$ has been presented by
\citet{suyu_1608_2}, in which they estimate a probability density
function for $\kappa_{\rm ext}$ using a method similar to that used in
\citet{hilbert_ms1}.  In particular, they trace rays through the
Millennium Simulation \citep{millennium}, and determine the
distributions of $\kappa_{\rm ext}$ between the redshift of the
background source and the observer.  However, rather than examining
the full distribution of $\kappa_{\rm ext}$ obtained from all lines of
sight, the $\kappa_{\rm ext}$ distribution is estimated using only
those lines of sight that have fractional galaxy number count
overdensities matching those observed in the field of the real lens.
The \citet{suyu_1608_2} analysis used our results for the fractional
overdensity along the B1608+656 line of sight ($\sim$2 compared to the
pure parallel sample) to obtain the prior on $\kappa_{\rm ext}$.  To
aid in similar analyses of time delay lens systems, we have provided
in Table~\ref{tab_nint} the integrated number counts and fractional
over/underdensities with respect to the pure parallel sample for each
of the lens systems considered here.

In order to explore the effects that line-of-sight variations in
$N_{\rm int}$ may have on future attempts to determine a global value
of $H_0$, we have undertaken a simple simulation.  This simulation
relies on four major assumptions, at least three of which are
questionable; the point of this exercise is solely to explore in a
very rough sense the possible implications of the number counts
measured in this paper.  The first assumption is that true
distribution of integrated number counts in lens fields is given by
the top panel of Fig.~\ref{fig_lens_vs_cosmos}.  Given our small
sample of only 20 lens systems, this assumption is almost certainly in
error.  We do eliminate the field with the highest integrated number
counts since that is associated with a clear cluster-scale lens and is
not representative of the galaxy-scale lenses in the other 19 fields.
The second assumption is that the true mean integrated number counts
are given by the values obtained from our analysis of the
pure-parallel sample.  The third assumption is that the overall
determination of $H_0$ from the lens sample is made without any
correction for variations in $\kappa_{\rm ext}$ from field to field.
The fourth assumption is discussed below.

\begin{figure}
\centering
\includegraphics[width=0.9\hsize]{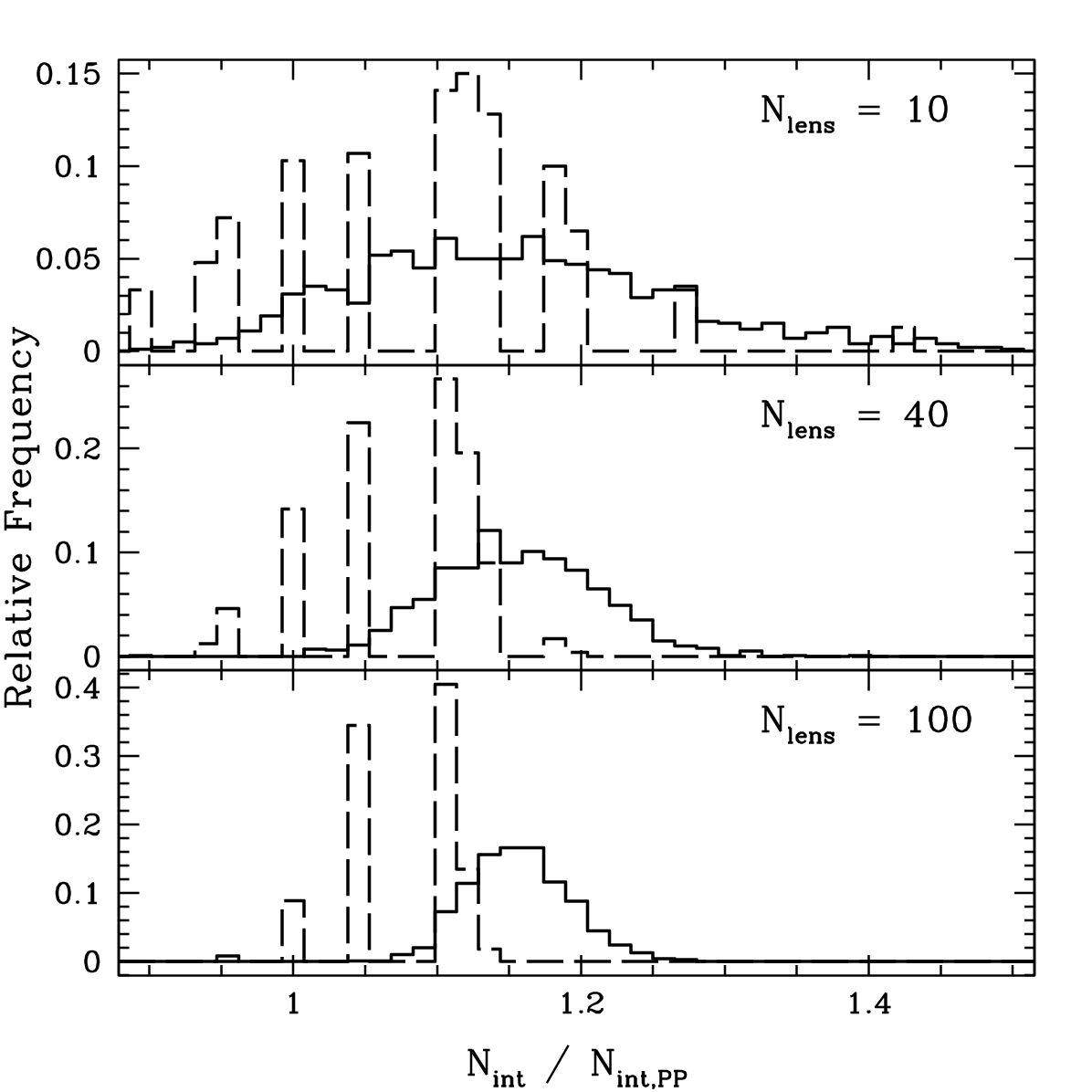}
\caption{Distributions of the sample means (solid curves) and medians (dashed
curves) of the integrated number counts obtained in simulated lens-field
samples.  In each case, 1000 realizations were obtained.  The three panels
show sample sizes of $N_{\rm lens} = 10, 40,$ and 100.
}
\label{fig_bootstrap}
\end{figure}

The first step in the simulation is to create 1000 realizations of a
sample of $N_{\rm lens}$ lenses via a bootstrap procedure, i.e., by
drawing $N_{\rm lens}$ values of $N_{\rm int}$ randomly, but with
replacement, from the distribution shown in the top panel of
Fig.~\ref{fig_lens_vs_cosmos}.  We consider three cases, $N_{\rm lens}
=$ 10, 40, and 100.  For each realization we calculate the mean and
median of the $N_{\rm int}$ distribution, and divide by $N_{\rm
int,PP}$ to facilitate our later estimates of $\kappa_{\rm ext}$.  The
resulting distributions are shown in Fig.~\ref{fig_bootstrap}.  The
widths of the distributions of mean values shows the expected $1 /
\sqrt{N}$ behaviour, with the sample RMS values being 0.11, 0.055, and
0.033 for $N_{\rm lens} = 10, 40$, and 100, respectively.  However,
our assumption that the true lens-field distribution of $N_{\rm int}$
is represented by the observed distribution given in
Table~\ref{tab_nint} and shown in Fig.~\ref{fig_lens_vs_cosmos}
introduces a bias into the lens field counts.  The centers of the
distributions occur at $N_{\rm int}/N_{\rm int,PP}$ of 1.16, 1.15, and
1.16 for the means and at 1.09, 1.08, and 1.08 for the medians.  Thus
the number counts in the simulated lens samples are biased high
compared to the pure-parallel sample by $\sim$15\% if a straight mean
is taken, and by $\sim$8\% if a median is taken, if no attempt is made
to correct for excess number counts in the lens fields.

The final step in the simulation is to convert the distributions of
$N_{\rm int}/N_{\rm int,PP}$ into distributions of $\kappa_{\rm ext}$.
This step involves the highly questionable final assumption, namely that
a simple conversion between $N_{\rm int}/N_{\rm int,PP}$ and $\kappa_{\rm ext}$
can be derived from the analysis of B1608+656 by \citet{suyu_1608_2}.
Their careful ray tracing through the Millennium Simulation found
that fields with $N_{\rm int}/N_{\rm int,PP} \sim 2$, corresponding to
the case for B1608+656, produced a distribution of $\kappa_{\rm ext}$
that was roughly centered at $\kappa_{\rm ext} \sim 0.1$.  One could assume
that this scaling held for all fields, and that $N_{\rm int,PP}$
corresponded to $\kappa_{\rm ext} \sim 0$.  Under these assumptions,
the bias in the lens-field integrated number counts of 
$N_{\rm int}/N_{\rm int,PP} \sim 1.15$ would correspond to 
$\kappa_{\rm ext} \sim 0.02$, and the resulting global determination of
$H_0$ from the lens sample would be biased high by $\sim$2\%.

However, the results of the simulation are based on the assumption
that no correction has been made for the estimated $\kappa_{\rm ext}$
in each lens field before averaging to obtain the global value of
$H_0$.  A lens for which a time delay has been measured (a crucial
component in determining $H_0$ from a lens system) will have been
imaged many times as part of a monitoring campaign.  Thus, the
integrated number counts in each of the lens fields should be known.
Therefore, it should be possible to reduce any bias due to improper
accounting for $\kappa_{\rm ext}$ in lens-based determinations of the
global value of $H_0$ if the following steps can be taken for each
lens in the sample.  (1) Quantitative estimates of the effect of the
local environment of the lens should be made.  These can come from
measurements of the velocity dispersion of the lensing galaxy
\citep[e.g.,][]{suyu_1608_2}, as well as photometric and/or
spectroscopic searches for a group or cluster that is physically
associated with the lensing galaxy.  (2) The galaxy number counts for
the lens field should be calculated within a reasonable aperture and
to a reasonable magnitude limit.  For the B1608+656 analysis,
\citet{suyu_1608_2} used the integrated counts ($19 \leq m \leq 24$)
within an aperture of radius 45\arcsec.  The magnitude limit should be
at or brighter than the completeness limit and should sample, as much
as possible, the full line of sight to the redshift of the background
object.  (3) Number counts with the same aperture and magnitude limits
should be computed for a control sample that is expected to be
representative of random lines of sight through the Universe.  For the
B1608+656 analysis, \citet{suyu_1608_2} used the pure-parallel sample
rather than the COSMOS sample, since the COSMOS number counts come
from a contiguous area on the sky that is not large enough to overcome
sample variance.  (4) After computing the ratio of the lens-field
number counts to the mean of the control sample counts, the
distribution of $\kappa_{\rm ext}$ can be estimated by ray tracing
through a structure formation simulation, such as the Millennium
Simulation \citep{millennium}.  Details on how this was done for the
B1608+656 field are given in \citet{suyu_1608_2}.  (5) This
$\kappa_{\rm ext}$ distribution should be used as a prior probability
in the full $H_0$ analysis for the lens system.

\begin{table}
\bigskip
\centering
\caption{Integrated number counts for all galaxies with $m \leq 24$
within 45\arcsec\ apertures.\label{tab_nint}}

\begin{tabular}{lrr}
\hline
Field & $N_{\rm int}$ & $[N_{\rm int}/N_{\rm int,PP}]$ \\
\hline
PP Mean         &   66 & 1.00 \\
COSMOS Mean     &   70 & 1.06 \\
JVAS B0218+357  &  125 & 1.89 \\
CLASS B0445+128 &   62 & 0.94 \\
CLASS B0850+054 &   75 & 1.14 \\
CLASS B1608+656 &  144 & 2.18 \\
CLASS B2108+213 &   74 & 1.12 \\
CFRS 03.1077    &   52 & 0.79 \\
HE 0435-1223    &   59 & 0.89 \\
HE 1113-0641    &   59 & 0.89 \\
J0743+1553      &   73 & 1.11 \\
J0816+5003      &   63 & 0.95 \\
J1004+1229      &   84 & 1.27 \\
RX J1131-1231   &   94 & 1.42 \\
SDSS 0246-0825  &   57 & 0.86 \\
SDSS 0903+5028  &   59 & 0.89 \\
SDSS 0924+0219  &   79 & 1.20 \\
SDSS 1004+4112  &  174 & 2.64 \\
SDSS 1138+0314  &   78 & 1.18 \\
SDSS 1155+6346  &   66 & 1.00 \\
SDSS 1226-0006  &   69 & 1.05 \\
WFI 2033-4723   &   88 & 1.33 \\
\hline
\end{tabular}
\end{table}

\section{Conclusions}

To assess whether strong gravitational lenses are preferentially found
along overdense lines of sight, we have devised a straightforward
Bayesian statistical number-count test which is conservative and
robust. It requires only the number counts of galaxies as function of
magnitude inside apertures of different sizes centered on the lenses
and on control fields. We have applied this method to a sample of 20 lenses with
F814W ACS images and control samples from the COSMOS and 
pure-parallel programs.  

Our hypothesis is that {\em if} gravitational lenses are found along
highly overdense lines of sight compared to random pointings on the
sky, either due to structure along the line of sight or overdensities
associated with the lens galaxies themselves, {\em then} the number of
galaxies within apertures centered on the lenses should show on
average more galaxies than similar apertures in the control
fields. This approach is conservative in that it does not require
redshifts for the galaxies (for a given magnitude, galaxies over
a wide redshift range contribute to the number counts), but only that
the ratio between mass and light integrated over the redshift cone is
close to constant. Thus, the number of galaxies can be used as a proxy
for mass; more galaxies on average implies more mass along the line of
sight.

More precise statistics can be constructed if the redshifts, galaxy
types, etc. are known, and the galaxy masses are derived through
scaling relations such as the Tully-Fisher relation
\citep{tully_fisher} or the fundamental plane
\citep{fp_dd,fp_dressler}.  However, the results of such models
quickly become model-dependent and prone to systematic errors.  Thus,
although such approaches may show a more significant result than ours,
a rejection of the hypothesis by our approach has the advantage of
being a robust result that is less dependent on model assumptions.

Having applied our number-count comparison to the selected lens and
control fields we find the following results:
\begin{enumerate}
 \item All distribution functions of number counts in the lens and
   control fields are well described by a combination of a Gaussian
   random field for the sample variance (i.e., the underlying density
   field varies from field to field) and Poisson statistics in the
   number counts. This defines our underlying Bayesian statistical
   model.
 \item In the three largest apertures, with radii of 45\arcsec,
   38\farcs1 and 22\farcs5 (steps of 2 in area), we find {\em no}
   significant difference in the number counts in the individual
   magnitude bins between the lens and
   control fields (Fig.~\ref{fig_delta_mu0} and
   Table~\ref{tab_mu0}). We emphasize that this does not presuppose
   that there is no overdensity, just that our robust statistics do
   not require it.  We also note that the uncertainties on the
   lens number counts are large, since there are only 20 lenses
   in our sample.
 \item The smallest aperture (15\farcs9) centered on the lenses,
   however, does show significantly more galaxies in the magnitude
   range $19 \leq m \leq 21$ than either the COSMOS or pure parallel
   fields. This is not unexpected since massive lens galaxies live in
   overdense regions (their two-point correlation is strong).
 \item When we correct for the effect of the two-point correlation,
   using the results from COSMOS by \citet{mccracken}, we find that a
   significant part of the differences between number counts in the
   lens and COSMOS fields can be accounted for by the fact that
   massive lens galaxies live in {\em locally} overdense regions, as
   expected.  The remaining differences are at the $\la
   1$--$\sigma$ level in the individual magnitude bins.
 \item Even though the differences are at the $\sim$1--$\sigma$ level
   for individual magnitude bins, the lens-field counts are
   consistently higher than the pure-parallel counts across several
   magnitude bins for the smallest and largest apertures.  This
   behaviour may indicate real overdensities, although once again the
   differences in the integrated number counts are only at the
   $\sim$1--2-$\sigma$ level.  These differences could either be due to
   small overdensities along the line of sight
   \citep[e.g.,][]{0712group, momcheva,1608groupdisc,auger1600}, or
   due to a too simplistic correction of the number counts for the
   COSMOS two-point correlation function.
 \item On average, the excess numbers of galaxies along the lines of
   sight to the lensing galaxies, integrated over the magnitude range
   that we have explored in this paper, are small once the correlation
   function corrections have been applied (e.g., $\la$4 galaxies
   for the largest aperture that we examined).  These excesses, which
   are mainly due to local overdensities associated with the lensing
   galaxy, amount to only $\sim$4--8\% of the total number counts in
   the three largest apertures.  
   These fractional
   overdensities can be compared with similar numbers derived from
   numerical simulations.  In particular, \citet{hilbert_ms1} found, by
   ray tracing through the Millennium Simulation \citep{millennium},
   that lenses do along biased lines of sight.  The contribution of the
   additional mass, however, was only a few percent of the total surface mass
   density along those lines of sight, in good agreement with our results.
 \item While the average number counts in our sample of 20 strong
   lenses does agree well with the control samples, once the effect of
   local clustering has been taken out, individual lens systems can
   still have significantly discrepant number counts.  Thus, 
   care should be taken when using an individual lens system to
   measure $H_0$.  We present a recipe for using number counts and
   other information to improve the treatment of line-of-sight convergence
   to measurements of $H_0$ using lenses.
\end{enumerate}

Based on our statistical test, we can say that: Yes, lens galaxies
do lie along overdense lines of sight compared to random pointings on
the sky, but these overdensities can be at least partially explained
by the fact that these massive lens galaxies are formed in locally
overdense regions.  We conclude that in strong-gravitational lens
modeling one always needs to assess the effect of the {\em local}
distribution of galaxies.  Our test indicates that the contribution by
everything else along the line of sight does not appear to be
significant on average.  However, a larger sample of lenses is needed
to strengthen any conclusions about the significance of differences
between the lens and control fields.  If the possible slight excess in
the lens field number counts persists in larger samples, then
global determinations of $H_0$ using lens samples may be biased unless
the analysis properly accounts for both the local overdensities
and the external convergence of the fields in which the lenses are
embedded.


\bigskip

We thank Phil Marshall, Tommaso Treu, and Maru\v{s}a Brada\v{c} for
fruitful discussions, and Phil for further insightful comments as the
referee of this paper.  We thank Phil Marshall and Tim Schrabback for
their work on the HAGGLeS pipeline and their help in modifying the
pipeline to run on the pure-parallel data.
CDF acknowledges support under HST program \#AR-10300, which was
provided by NASA through a grant from the Space Telescope Science
Institute, which is operated by the Association of Universities for
Research in Astronomy, Inc., under NASA contract NAS 5-26555.  
He is also grateful for the generous hospitality shown by the Kapteyn
Institute on his visits there.
LVEK is supported (in part) through an NWO-VIDI program subsidy
(project number 639.042.505).
This work is supported by the European Community's Sixth Framework Marie 
Curie Research Training Network Programme, Contract No. 
MRTN-CT-2004-505183 `ANGLES'.



\label{lastpage}

\end{document}